# Monolithic CMOS-compatible zero-index metamaterials

Daryl I. Vulis[†], Yang Li[†], Orad Reshef[†], Philip Camayd-Muñoz, Mei Yin, Shota Kita, Marko Lončar, Eric Mazur

*Department of Physics and Division of Engineering and Applied Sciences, Harvard University, 9 Oxford Street, Cambridge, Massachusetts 02138*

[†] These authors contributed equally to this work.

Abstract: Zero-index materials exhibit exotic optical properties that can be utilized for integrated-optics applications. However, practical implementation requires compatibility with complementary metallic-oxide-semiconductor (CMOS) technologies. We demonstrate a CMOS-compatible zero-index metamaterial consisting of a square array of air holes in a 220-nm-thick silicon-on-insulator (SOI) wafer. This design is achieved through a Dirac-cone dispersion. The metamaterial is entirely composed of silicon and offers compatibility through low-aspect-ratio structures that can be simply fabricated in a standard device layer. This platform enables mass adoption and exploration of zero-index-based photonic devices at low cost and high fidelity.

**Introduction**

Silicon photonics has become the principal platform for on-chip optical telecommunications [1–3]. Furthermore, monolithic fabrication of many silicon-based photonic devices allows for integration with existing CMOS technologies that currently dominate computation and communications [3,4]. As integrated silicon photonics approaches the nanoscale, solutions based on new materials are necessary to manage the confinement and behavior of light. Metamaterials offer a way to engineer materials with novel, tailored properties that extend the capabilities of silicon photonics.

Zero-refractive-index metamaterials have attracted attention for engineering the interaction of light in integrated photonics [5–12]. The effective wavelength within this material is infinite, allowing phase-free propagation of light as well as strong nonlinear optical enhancement [13–17]. Such extreme properties have inspired a variety of applications, including efficient coupling between disparate modes and sub-diffraction limit waveguide design [8,18,19], electromagnetic cloaking [20,21], beam steering [22], quantum optics [23,24], and new approaches to phase-matching in nonlinear optics [17,25,26].

Effective implementation of a zero-refractive-index metamaterial in integrated photonics requires both CMOS compatibility and monolithic fabrication: the zero-index metamaterial platform must 1) be fully dielectric, 2) contain low-aspect-ratio structures for compatibility with standard SOI wafers, and 3) operate at telecom frequencies. Integrated applications also necessitate an in-plane geometry in which light propagates parallel to the substrate surface for compatibility with silicon waveguides or photonic devices.

Inspection of the definition of refractive index ($n^{\text{eff}} = \sqrt{\varepsilon^{\text{eff}} \mu^{\text{eff}}}$) reveals that an index of zero is achieved if either or both the electric ($\varepsilon^{\text{eff}}$) or magnetic ($\mu^{\text{eff}}$) response of a material is tuned to zero. While we can tune the electric and magnetic response using metallic inclusions [8,27], incorporating metals results in Ohmic losses and incompatibility with all-dielectric silicon photonics. In addition, inspection of the impedance equation ($Z^{\text{eff}} = \sqrt{\mu^{\text{eff}}/\varepsilon^{\text{eff}}}$) reveals that this "single zero" case produces an impedance of either infinity or zero — a mismatch that makes such metamaterials unsuitable for integrated applications. Fully dielectric zero-index metamaterials present an attractive alternative. Such metamaterials can be achieved by tuning Mie resonances in a photonic crystal structure [6,10]. This approach allow for precise and simultaneous control of the electric and magnetic response and can be used to tune both the effective relative permittivity $\varepsilon^{eff}$ and effective relative permeability $\mu^{eff}$ to zero simultaneously, yielding a finite impedance [6,7].

Previous all-dielectric zero-index metamaterial designs are largely unsuitable for integrated photonic applications. For example, out-of-plane configurations in which light propagates perpendicular to the substrate surface are impractical for integrated applications or long interaction lengths [7]. In comparison, in-plane configurations formed by cascading regions of both negative and positive phase advance do not offer flexible shape configuration or isotropic behavior [9]. Finally, zero-refractive-index metamaterials have been demonstrated by our group for an in-plane configuration with isotropic behavior. However, these metamaterials rely on either a unit cell of artificially infinitely tall silicon pillars achieved through parallel conductors [28] or silicon pillars whose height considerably exceeds current standard SOI platforms [29].

Here, we present the first monolithically-fabricable, CMOS-compatible Dirac-cone metamaterial with an impedance-matched refractive index of zero for telecom frequencies. This metamaterial and other photonic devices can be patterned and structured simultaneously on a photonic chip based on a standard material platform of 220-nm-thick SOI (Fig. 1). In addition, this monolithic, all-dielectric platform is the first experimental demonstration of a zero-index metamaterial through the degeneracy of dipole and quadrupole modes.

The metamaterial design consists of a square array of air holes in 220-nm-thick SOI, which can be patterned using a single electron-beam-lithography procedure. The air holes are then structured through an inductively coupled plasma reactive ion etch procedure. To measure the index of refraction, we construct a triangular prism of this metamaterial and observe that the transverse-electric (TE) polarized light entering the prism is refracted perpendicular to the output facet through a semi-circular polymer waveguide. A zero degree angle of refraction, combined with a linear dispersion, unambiguously demonstrates a refractive index of zero.

Through configuration of the unit cells, the metamaterial can be defined in arbitrary shapes for varied potential applications. In addition, use of standard planar fabrication techniques maintains intrinsic compatibility with the vast library of existing SOI-based devices. This metamaterial enables the mass production of zero-index-based devices at low cost and high fidelity through CMOS fabrication techniques and simple features consisting of nanoscale circular holes. Combined with a monolithic fabrication procedure, this platform provides an easily adaptable method of exploring various applications of a refractive index near zero.

**Design and Theory**

To achieve a platform with low-aspect-ratio features, we increase the ratio of silicon to air as compared to previous zero-index designs based on infinitely tall silicon pillars [6,8,10,28]. An air hole in silicon matrix design increases the volume filling fraction of silicon, allowing the area between defined air columns to be used as short resonators suitable for thinner silicon films. The metamaterial is designed for transverse electric (TE) polarized modes that are more strongly confined in thin silicon layers than transverse magnetic (TM) polarized modes [30,31].

Furthermore, we expect symmetric modes to result from the symmetric unit cell lattice featured in an air hole structure; such modes can form a degeneracy corresponding to a Dirac cone dispersion [32]. We tune the size and separation of the air holes in our metamaterial to ensure the degeneracy of three symmetric modes at the gamma point and achieve a photonic Dirac cone at the center of the Brillouin zone (Fig. 2b) for a wavelength of 1550 nm. The pitch and radius are 738 nm and 222 nm, respectively. Despite the low-aspect ratio, this metamaterial maintains propagation losses that are comparable to previously demonstrated zero-index metamaterials [28].

The calculated bandstructure for the designed material is shown in Fig. 2a. Two bands with linear dispersion intersect at the $\Gamma$-point, forming a Dirac-like cone [6,10]. We also observe an additional "flat" band corresponding to a longitudinal electric dipole mode. The bands display quadratic dispersion in the vicinity of the $\Gamma$-point, deviating from an ideal (linear) Dirac cone. The quadratic bands cause a deformation in the linear Dirac-cone dispersion with characteristics similar to previously proposed exceptional points or rings [33]. The shape of this dispersion is more clearly observed in the dispersion surface shown in Fig. 2b. For a sufficiently small exceptional ring diameter, we achieve a degeneracy of the modes at the $\Gamma$-point and expect behavior similar to that caused by a linear Dirac-cone dispersion.

This work represents the first demonstration of zero-index via degeneracy involving a quadrupole mode. The intersection of the two Dirac-cone bands corresponds to a degeneracy of electric dipole and magnetic quadrupole modes at the $\Gamma$-point (Fig. 2c). While Dirac-cone or exceptional-ring behavior as the product of dipole and quadrupole modes has been predicted [22,32,33], treatment of a material with a Dirac-cone dispersion comprised of these modes as an effectively zero-index medium may be challenging and has not yet been demonstrated unambiguously [34]. This necessitates special consideration of homogenization criterion to ensure that the modes excited in our metamaterial by a normal incidence plane-wave source produce an effectively zero-index behavior.

To visualize the photon lifetime, the quality factor of the modes is superimposed on the band structure in Figure 2a. The observed difference in quality factor between the dipole and quadrupole modes, estimated at 40 and 5500 respectively, causes the quadratic dispersion near the Γ-point. Radiative losses produced by the dipole mode disrupts the degenerate behavior and results in a lower group velocity near the Γ-point, producing the quadratic bands in that region [33,35]. Conversely, the quadrupole modes do not produce significant radiation losses as there is a symmetry mismatch of that mode in relation to plane waves (Fig. 2c).

We extract the effective refractive index and impedance of the metamaterial using simulated transmission and reflection coefficients (Fig. 2d) and effective medium theory [36]. We find that the refractive index crosses zero with linear dispersion in the vicinity of the Dirac cone at the design wavelength of 1550 nm. We also see that the material has an effective normalized impedance of 0.8. From the surface in Fig. 2b and the collapsed isofrequency contours corresponding to frequencies above and below the design wavelength in Fig. 2e, we see that the modes at the Γ-point form nearly circular isofrequency contours close to the design wavelength, indicating isotropic behavior.

**Experimental results**

To experimentally measure the refractive index, we constructed a right triangular prism of this metamaterial and measured the refraction of TE-polarized light entering the prism. Figure 3 shows a scanning electron microscope (SEM) image of the fabricated device, including input and output coupling structures (Fig. 3a), and a close-up image of the fabricated metamaterial prism (Fig. 3b). TE-polarized light is coupled into the metamaterial through a tapered silicon waveguide. This ensures plane-wave excitation of the modes corresponding to a refractive index of zero. Light exits the prism via the output facet into a semicircular polymer (SU-8) slab waveguide. The exiting light propagates via the slab waveguide. At the end of the polymer waveguide, scattering is enhanced by a silicon lip under the outside edge (Fig. 3a). To extract the angle of refraction, $\alpha$, the scattered beam is imaged by a near-infrared (NIR) camera (Fig. 3b).

We measure the angle of refraction of the metamaterial by sampling the intensity at the outer edge of the SU-8 slab waveguide for each of the wavelengths ranging from 1480 nm through 1680 nm (Fig. 4a). The data show a clear linear transition from a positive-index regime to a negative-index regime, crossing zero at a wavelength of 1625 nm. We also observe an additional, weaker beam at +45°. We attribute this additional beam to TM modes in the metamaterial, as we are unable to completely eliminate TM polarized light from the input source. This conclusion is supported by TM far-field projections (FFP). We also simulated the refraction through a prism with the same design parameters as the fabricated prism (as measured by SEM). As Fig. 4b shows, the experimental results are in excellent agreement with the simulation and the two exhibit the same zero-crossing wavelength (Fig. 4b). The additional lobes in the simulated FFP at approximately +20° and −45° for wavelengths near 1625 nm are produced by design parameter deviations in the simulated prism.

Figure 4b shows a representative image of the refraction through the prism at a wavelength of 1625 nm, including the scattering from the edge of the slab waveguide. The refracted beam is observed directly above the prism, indicating refraction normal to the output facet and an effective index of zero. This shift in zero-index wavelength from the design target of 1550 nm can be attributed to fabrication imperfections.

We can extract the effective refractive index $n^{\text{eff}}$ of the metamaterial prism from Snell's law,

$$n_{\text{SU8}}/n^{\text{eff}} = \sin 45° / \sin \alpha ,$$

where the angle of incidence on the output facet of the prism has been taken to be 45°, $n_{\text{SU8}} = 1.58$ is the index of the output polymer slab waveguide, and $\alpha$ is the measured angle of the scattered beam (Fig. 3b). Figure 4d shows the index of refraction obtained this way from both the measurements and the simulation over the wavelength range from 1480 nm to 1680 nm. The extracted index ranges from 0.51 ± 0.04 at 1480 nm to −0.21 ± 0.05 at 1680 nm. The error bars in the experimental data are due to finite image resolution and fitting uncertainty. The fitting procedure is described in detail in previous work [28]. The relationship between index and wavelength is linear along the entire measurement range from 1480 nm through 1680 nm and in the vicinity of the zero-index wavelength of 1625 nm.

**Robustness of modal degeneracy against fabrication imperfections**

To consider the robustness of the modal degeneracy at the Γ-point against fabrication imperfections, we simulate prisms with larger and smaller air-hole radii. Within a range of 0.8% of the target radii and 0.3% of the target pitch we find that the zero-crossing wavelength redshifts or blueshifts, respectively. Most importantly, slight fabrication imperfections still produce the desired modal degeneracy and resultant zero-index behavior. At larger deviations from the ideal design, however, we observe the formation of photonic bandgaps. The appearance of these bandgaps is a further indication that the experimentally observed linear-dispersion behavior near the zero-crossing is the Dirac-cone-like dispersion.

**Conclusion**

We presented the first CMOS-compatible zero-index metamaterial in the optical regime, featuring monolithic fabrication on an industry-standard 220-nm-thick SOI wafer. By eliminating metallic inclusions and high-aspect-ratio structures, this zero-index platform offers a significantly simplified fabrication procedure over earlier designs and compatibility with silicon foundries. In addition, satisfaction of homogenization criterion for normal incident light permits treatment of the platform as a metamaterial. As such, this metamaterial can be implemented in several proposed zero-index applications, including supercoupling [18], phase matching in nonlinear optics [25], and beam steering [22]. This design methodology can be extended to achieve a refractive index of zero using even thinner SOI layers and different material systems. In addition, this is the first experimental demonstration of an exceptional ring-based zero-index metamaterial formed by the degeneracy of dipole and quadrupole modes. Integration with current silicon photonic platforms through monolithic fabrication and impedance matching offers a powerful platform for exploring the applications of zero-index materials.

**Methods**

<u>Simulation</u>

The complex indices of silicon and silica for use in all numerical simulations were measured using spectroscopic ellipsometry. The complex index and material parameters, and the far-field patterns were calculated using a commercial three-dimensional finite-difference time-domain solver. Reflected and transmitted electric fields were extracted at one point before the source and one after the metamaterial, respectively, to obtain the complex reflection and transmission coefficients. The electric fields were collected from a near-field simulation to obtain the far-field pattern using far-field projection.

The band structures, electromagnetic mode profiles, dispersion surfaces and isofrequency contours were computed using a commercial three-dimensional finite element method solver. These results were obtained by calculating all the modes in a unit cell of the metamaterial with Floquet periodic boundary conditions in the *x* and *y* directions and perfectly matched layers at the boundaries in the *z* direction (*i.e.,* out of the plane). TE-polarized modes were selected by evaluating the energy ratio of the magnetic fields in the *x*, *y* and *z* directions. Modes with low quality factors (<20) were filtered out.

Fabrication

The metamaterial is fabricated using an SOI wafer with a 220-nm-thick silicon top layer. First, the air-hole array, input silicon waveguides, and silicon lip that is beneath the outside edge of the SU-8 slab waveguide are patterned via electron-beam lithography (EBL) into negative-tone resist (XR-1541 6%, Dow Corning). Reactive ion etching in an $SF_6:C_4F_8$ atmosphere is then used to remove the silicon in the air holes and silicon surrounding the waveguides as shown in Figure 1. The remaining negative-tone resist is then removed using 7:1 Buffered Oxide Etch (BOE). A 1.5-µm-thick SU-8 layer is spin-coated and patterned using EBL to form the output slab waveguide, then cured. Finally, a 2-μm-thick SU-8 waveguide is fabricated using the same method to form the calibration waveguide used to align images taken with the NIR camera.

**Acknowledgements**

The authors thank Chua Song Liang, Zin Lin, Olivia Mello, Cleaven Chia, Haoning Tang, Dario Rosenstock, and Lysander Christakis for helpful discussions and assistance with simulations. The research described in this Article was supported by the National Science Foundation under contract DMR-1360889, the Air Force Office of Scientific Research under contract FA9550-14-1-0389, the National Science Foundation Graduate Research Fellowship Program, the National Department Science and Engineering Graduate Fellowship, and the Natural Sciences and Engineering Research Council of Canada. This work was performed in part at the Center for Nanoscale Systems (CNS), a member of the National Nanotechnology Coordinated Infrastructure Network (NNCI), which is supported by the National Science Foundation under NSF award no. 1541959. CNS is part of Harvard University.

**Author contributions**

Y.L. conceived the basic idea for this work. Y.L., O.R., P.M., and D.V. carried out the FDTD simulations. O.R., S.K, P.M., and M.Y. carried out the FEM simulations. O.R. and D.V. designed the optical waveguides. D.V. performed the fabrication. O.R. and D.V. carried out the measurements. P.M. analyzed the experimental results. M.L. and E.M. supervised the research and the development of the manuscript. D.V. wrote the manuscript, and all authors subsequently took part in the revision process and approved the final copy of the manuscript.

## Figures

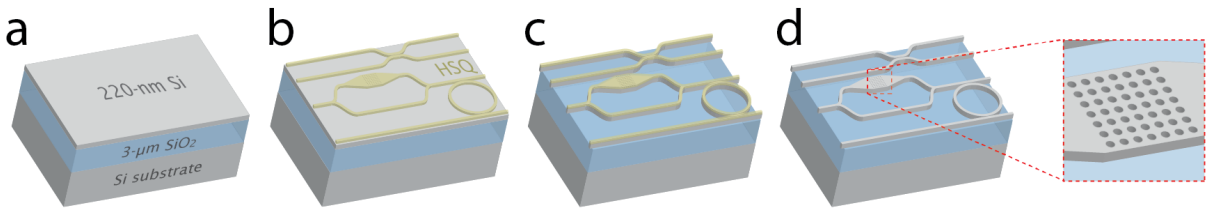

**Figure 1: Platform and fabrication.** Schematic showing the fabrication stages of the metamaterial and its intrinsic compatibility with silicon photonics. The fabrication of all elements could be done simultaneously in a single, monolithic procedure as illustrated above. a) A substrate consisting of 220-nm-thick SOI is b) patterned via electron beam lithography, then c) structured through RIE-ICP and finally d) the resist is removed revealing the completed structure. **Inset:** Close-up view of a metamaterial air-hole array.

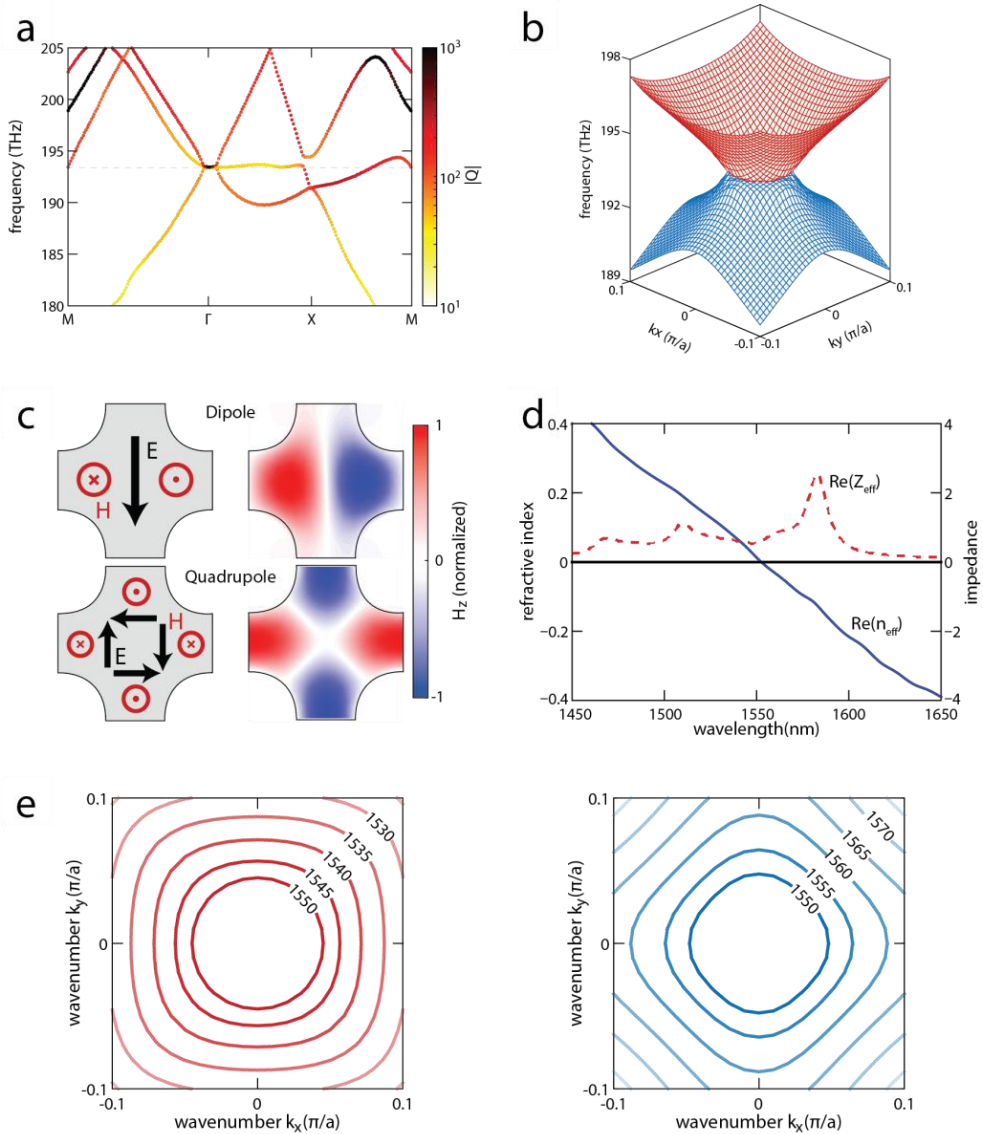

**Figure 2: Design of the zero-index metamaterial. a)** Photonic band structure of the zero-index metamaterial for TE modes. Two linear dispersion bands intersect at the Γ-point at λ = 1550 nm. Quality factor of the modes is superimposed on band structure showing strong confinement of modes forming the degeneracy. Disparity between the quality factors of the two bands produces a quadratic dispersion in the region of the Γ-point. **b)** Three-dimensional dispersion surfaces. The linear bands (blue and red) form a Dirac-like cone. We show only the two modes that form the cone to emphasize the Dirac-cone dispersion. **c)** Electric fields at the Γ-point over a unit-cell cross-section in the plane of the array, corresponding to an electric dipole mode and a magnetic quadrupole mode. **d)** Effective index (blue line) and normalized impedance (red line) of the metamaterial retrieved from simulated reflection and transmission coefficients. The index crosses zero linearly at a wavelength of 1550 nm. Impedance is shown to have a finite value of 0.6 at the design wavelength. **e)** Isowavelength contours of the zero-index metamaterial for wavelengths below (left) and above (right) the design wavelength. Lighter shades indicate wavelengths farther from the design wavelength. The nearly circular contours indicate that this metamaterial is almost isotropic near the Γ point.

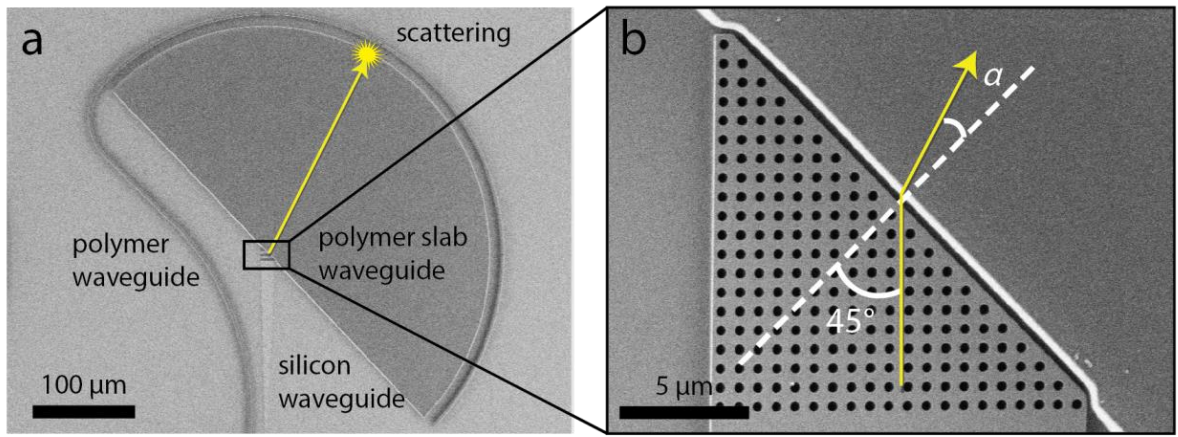

**Figure 3: Fabricated metamaterial and experimental setup. a)** Scanning electron microscopy image of the fabricated device. A silicon waveguide carries the incident beam towards the zero-index metamaterial prism as outlined in black, where the beam is refracted into the SU-8 slab waveguide. A silicon lip at the outside edge of the slab waveguide is used to scatter the output beam for optimal imaging. The angle of refraction $α$ is determined by measuring the position of the refracted beam at the curved output edge of polymer slab waveguide as indicated by the yellow scattering point. An additional polymer waveguide around the outside edge of the slab waveguide includes defects that are used to align the infrared images during experimental data processing. **b)** Fabricated zero-index metamaterial prism showing the incident and refracted beams.

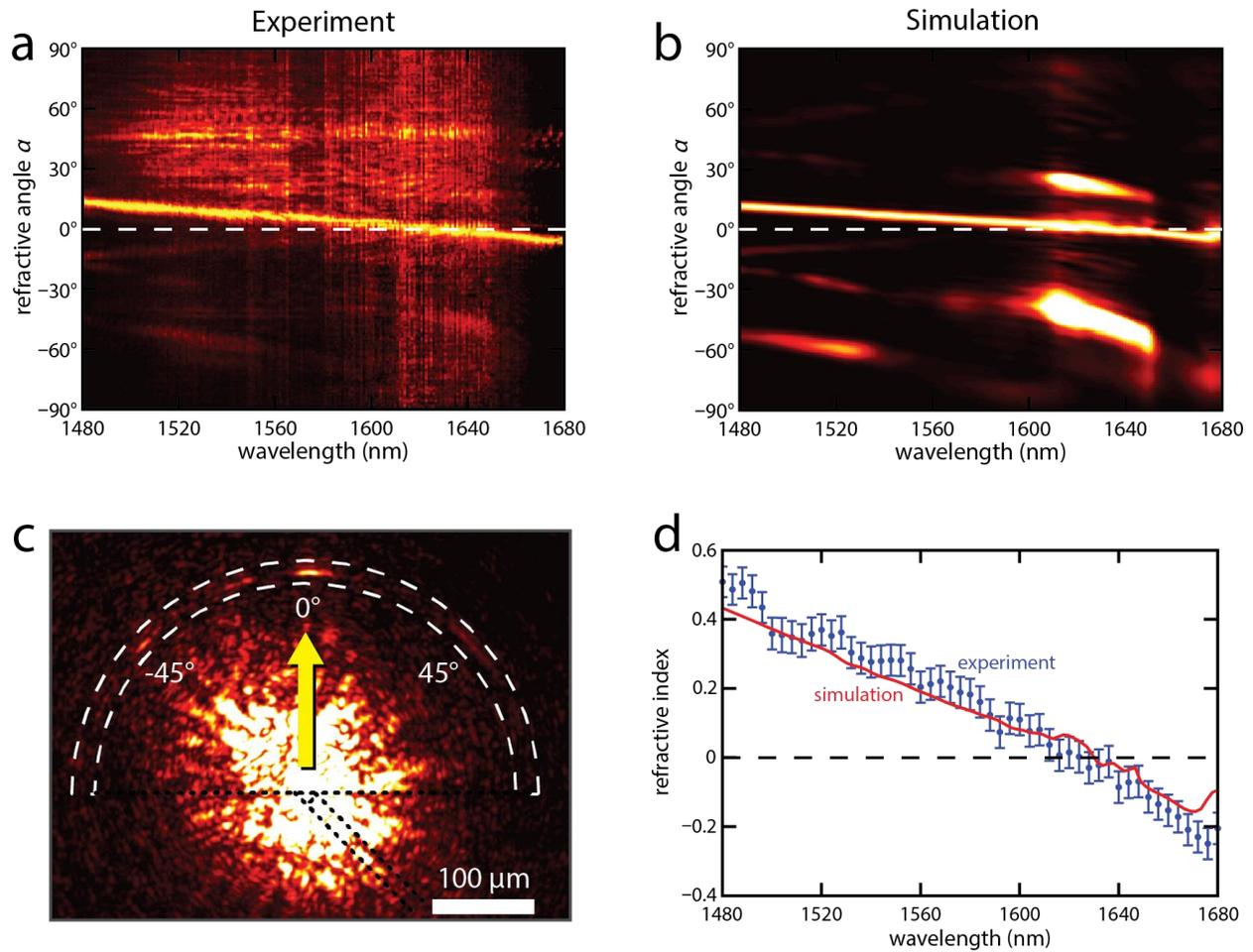

**Figure 4: Experimental refractive index measurement. a)** Measured near-field pattern and **b)** corresponding simulated far-field pattern (right). The white dashed lines show that the refracted beam crosses 0° at a wavelength of 1625 nm. The image is normalized to the maximum intensity at each wavelength. **c)** Near-infrared microscope image of the prism (Fig. 3) at 1625 nm, showing the refracted beam, which propagates normal to the interface between the prism and the SU-8 slab waveguide. The black dotted lines indicate the position of the prism and input waveguide. The white dashed lines delineate the portion of the image that is used to produce the measured near-field pattern (Fig. 4a). **d)** Effective index of the metamaterial extracted from the measured (blue dots) and simulated (red line) angles of refraction, $\alpha$.